\documentclass{eas}
\usepackage{graphicx}
\usepackage{array}
\usepackage{amsmath}
\usepackage{graphicx}
\usepackage{amssymb}
\begin{document}

\title{Dark Energy-Dark
Matter Interaction from the Abell Cluster A586}
\runningtitle{DE-DM interaction from the Abell cluster A586}

\author{Orfeu Bertolami}\address{\textbf{Speaker}; Departamento de F\'{\i}sica, Instituto Superior T\'{e}cnico,  Av. Rovisco Pais 1, 1049-001 Lisboa, Portugal. E-mail: orfeu@cosmos.ist.utl.pt}
\author{Francisco Gil Pedro}\address{Departamento de F\'{\i}sica, Instituto Superior T\'{e}cnico,  Av. Rovisco Pais 1, 1049-001 Lisboa, Portugal. E-mail: fgpedro@fisica.ist.utl.pt}
\author {Morgan Le Delliou}\address{Centro de F\'{\i}sica Te\'{o}rica e Computacional, Universidade de Lisboa, Av. Gama Pinto 2, 1649-003 Lisboa, Portugal. E-mail: delliou@cii.fc.ul.pt}

\begin{abstract}
	We find that deviation from the virial equilibrium of the Abell Cluster A586 yields evidence 			of the interaction	between dark matter and dark energy. We argue that this interaction
	might imply a violation of the Equivalence Principle. These evidences are found in the context of two different models of dark	energy-dark matter interaction.
\end{abstract}
\maketitle
\section{Introduction}

Current cosmological data strongly suggests that the existence dark energy (DE) and dark matter (DM) is crucial for a suitable description of universe's evolution. 

Even though observations
are consistent with the $\Lambda$CDM model, a deeper insight into the nature of DE and DM
might require more complex models -- in particular considering the interaction between
these components. However, so far, no evidence of this putative interaction has been presented.
In this work, we argue that data on the cluster A586 shows evidence that DE-DM interaction is an active dynamical factor. Furthermore, we argue that this suggests evidence of violation of the Equivalence
Principle (EP).

In what follows, we set the general framework to treat the interaction between DE and DM
and consider two very different, phenomenologically viable models:
one based on \textit{ad hoc} DE-DM interaction [\cite{Amendola}], the
other on the generalized Chaplygin gas (GCG) model with explicit identification
of DE and DM [\cite{Bento04}]. Our observational inferences are based
on cluster A586, given its stationarity, spherical symmetry and wealth
of available observations [\cite{Cypriano:2005}]. We also compare our
results with other cosmological observations [\cite{Guo07}].

\section{DE-DM Interacting models}

Our results are obtained in the context of two distinct phenomenologically
viable models for the DE-DM interaction: the DE-DM unification model,
the GCG [\cite{Bento02}], but also a less constrained interacting model
with constant DE equation of state (hereafter EOS) parameter $\omega_{DE}=p_{DE}/\rho_{DE}$ (see e.g. \cite{Amendola}).

We consider first a quintessence model with constant EOS. The Bianchi
conservation equations for both DE and DM read -- $H$ denoting the
Hubble parameter and $\zeta$, the interaction strength:
\begin{center}\begin{eqnarray}
\dot{\rho}_{DM}+3H\rho_{DM} & = & \zeta H\rho_{DM},\\
\dot{\rho}_{DE}+3H\rho_{DE}(1+\omega_{DE}) & = & -\zeta H\rho_{DM},\end{eqnarray}
\end{center}

\noindent for a constant EOS parameter ($\omega_{DE}$), and scaling ($\eta$) $\rho_{DE}/\rho_{DM}=\Omega_{DE_{0}} a^{\eta}/\Omega_{DM_{0}}$
are assumed. Thus, it follows that the coupling varies as

\begin{equation}
	\zeta=-\frac{(\eta+3\omega_{DE})\Omega_{DE_{0}}}{\Omega_{DE_{0}}+\Omega_{DM_{0}}a^{-\eta}}
\end{equation}

and

\begin{align}
\rho_{{\scriptscriptstyle DM}}= & a^{-3}\rho_{{\scriptscriptstyle DM_{0}}}\left[\Omega_{{\scriptscriptstyle DE_{0}}}a^{\eta}+\Omega_{{\scriptscriptstyle DM_{0}}}\right]^{-\frac{(\eta+3\omega_{{\scriptscriptstyle DE}})}{\eta}}, & \rho_{{\scriptscriptstyle DE}}= & a^{\eta}\rho_{{\scriptscriptstyle DE_{0}}}\frac{\rho_{{\scriptscriptstyle DM}}}{\rho_{{\scriptscriptstyle DM_{0}}}}.
\end{align}

We turn now to the GCG model, which is defined by its unified EOS $p_{DE}=-A/(\rho_{DM}+\rho_{DE})^{\alpha}$,
with $0<A\leq 1$ and $0\leq\alpha\leq 1$, and assuming DE constant EOS parameter
$\omega_{DE}=-1$ [\cite{Bento04}]. The splitting into DE and
DM, discussed in \cite{Bento04} together with Bianchi conservation imply a scaling behaviour with
$\eta=3(1+\alpha)$ and energy densities to be

\begin{center}\begin{align}
\rho_{DM} & =\rho_{DM_{0}}\frac{\rho_{DE}}{\rho_{DE_{0}}}a^{-3(1+\alpha)}, & \rho_{DE}= & \rho_{DE_{0}}\left(\frac{\Omega_{DE_{0}}+\Omega_{DM_{0}}}{\Omega_{DE_{0}}+\Omega_{DM_{0}}a^{-3(1+\alpha)}}\right)^{\frac{\alpha}{1+\alpha}}.\end{align}
\end{center}

\section{Generalized Layzer-Irvine equations}

We focus, as described in [\cite{Berto07}a], on the effect of interaction on clustering
as revealed by the Layzer-Irvine equation. We write the kinetic and
potential energy densities $\rho_{K}$ and $\rho_{W}$ of clustering
DM considering the interaction with DE in terms of scale factor dependence

\noindent \begin{align}
\rho_{K} & \propto a^{-2}, & \rho_{W} & \propto a^{\zeta-1},\end{align}

\noindent and we use DM virialization dynamics in an expanding universe described
by the generalised Layzer-Irvine equation [\cite{Peebles}] to obtain [\cite{Berto07}a]:

\noindent \begin{center}\begin{align}
\dot{\rho}_{DM}+\left(2\rho_{K}+\rho_{W}\right)H & =\zeta H\rho_{W},\end{align}
\end{center}

For the GCG, the scaling allows to just replace $\eta=3(1+\alpha)$ and $\omega_{DE}=-1$.
At equilibrium, in the presence of interaction, $2\rho_{K}+\rho_{W}=\zeta\rho_{W}\ne0$.

\section{Detection of interaction from the observation of A586}

Using those models, we have considered observations from the Abell
cluster A586 and compared it with other observations [\cite{Berto07}a,b].

\begin{figure}
\begin{center}\includegraphics[%
  width=0.35\paperwidth,
  height=0.67\textheight,
  keepaspectratio]{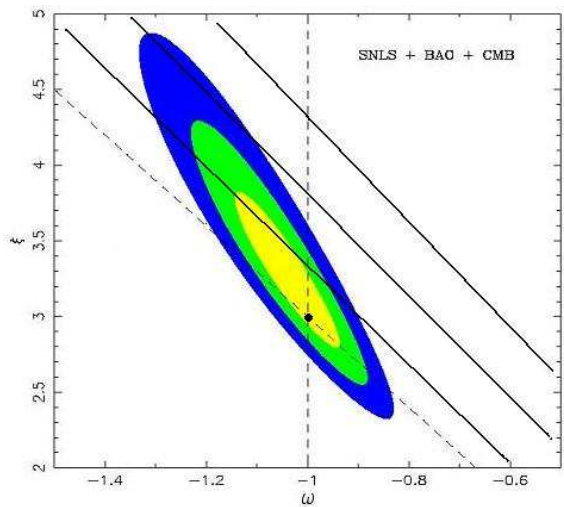}
\end{center}
\caption{\label{fig:GuoPlus}Superimposition of the probability contours for
the interacting DE-DM model in the [$\omega_{X}$,$\xi$] plane (denoted
as ($\omega_{DE_{0}}$,$\eta$) in [\cite{Berto07}a,b]), marginalized
over $\Omega_{DE_{0}}$ for CMB, SN-Ia and BAO
(2.66$<\xi<$4.05 at 95\% C.L.) observations [\cite{Guo07}] with the results extended from [\cite{Berto07}a,b],
based on the study of A586 cluster. The $\xi=-3\omega_{X}$ line corresponds
to uncoupled models.}
\end{figure}

From \cite{Cypriano:2005}, we extract: the total mass, $M_{Cluster}=(4.3\pm0.7)\times10^{14}\: M_{\odot}$
(galaxies, DM and intra-cluster gas), radius, $R_{Cluster}=$ 422
kpc at $z=0.1708$ (angular radius $\Delta_{max}=145''$) and cluster
velocity dispersion $\sigma_{v}=(1243\pm58)\: kms^{-1}$, from weak lensing.
From photometry, we can obtain the mean intergalactic distance:
\begin{center}
 $<R>=\frac{2}{N_{gal}(N_{gal}-1)}\sum_{i=2}^{N_{gal}}{\sum_{j=1}^{i-1}r_{ij}}$,\\	
\end{center}
where $r_{ij}^{2}=2d^{2}\left[1-\cos(\alpha_{ci}-\alpha_{cj})\cos\delta_{ci}\cos\delta_{cj}-\sin\delta_{ci}\sin\delta_{cj}\right]$,
$\alpha_{ci}=\alpha_{i}-\alpha_{center}$ and $\delta_{ci}=\delta_{i}-\delta_{center}$%
, from declinations and right assentions of a galaxy sub-sample within
the projected $\Delta_{max}$ where for galaxy $i$, $\sqrt{\alpha_{ci}^{2}+\delta_{ci}^{2}}\leq\Delta_{max}$.\enlargethispage{.25cm}
Further assuming $\omega_{DE}=-1$ and $\Omega_{DE_{0}}=0.72$, $\Omega_{DM_{0}}=0.24$
[\cite{Spergel:2006hy}], we get 

\begin{align}
\rho_{K} & \simeq\frac{9}{8\pi}\frac{M_{Cluster}}{R_{Cluster}^{3}}\sigma_{v}^{2}=(2.14\pm0.55)\times10^{-10}Jm^{-3},\\
\rho_{W} & \simeq-\frac{3}{8\pi}\frac{G}{<R>}\frac{M_{Cluster}^{2}}{R_{Cluster}^{3}}=(-2.83\pm0.92)\times10^{-10}Jm^{-3},
\end{align}

\noindent which allows one to obtain
\begin{align}
\eta & =3.82_{-0.47}^{+0.5}, & \alpha & =0.27_{-0.16}^{+0.17}.
\end{align}

Notice that $\eta\neq-3\omega_{DE}$ signals the energy exchange between
DM and DE. It is also remarkable that $\alpha\neq0$ which implies
the GCG description is not degenerate with $\Lambda$CDM ($\alpha=0$).
We mention that our results (see Fig.\ref{fig:GuoPlus} and \cite{OBPedro07b})
are consistent with the study of \cite{Guo07} (see also \cite{Amendola}) where DE-DM interacting
quintessence are analysed for compatibility with WMAP CMB [\cite{Spergel:2006hy}],
SNLS SN-Ia [\cite{SNLS06}] and Baryon Acoustic Oscillations in SDSS
[\cite{BAOinSDSS}].

\section{Putative Violation of the Equivalence Principle}

Given that the EP concerns the way matter falls in the gravitational
field, considering the clustering of matter against the cosmic expansion
and the interaction with DE seems to be a logical way to test the
validity of this fundamental principle. Both models \emph{predict} departure of homogeneous DM from
dust behaviour and have effects that can be interpreted as violation
of EP.%

\begin{figure}[ht]
	\begin{center}\includegraphics[%
	  width=0.35\paperwidth,
	  height=0.67\textheight,
	  keepaspectratio]{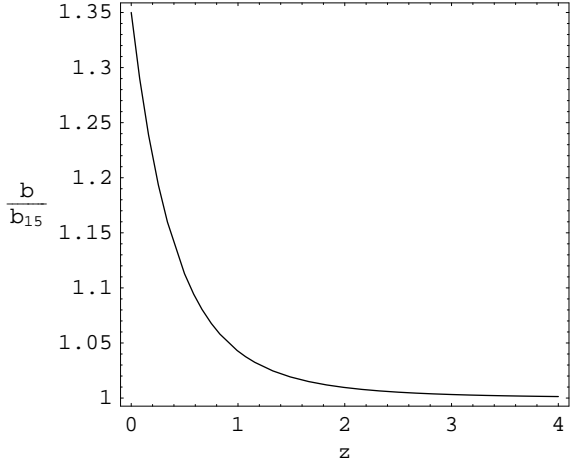}
	\end{center}

	\caption{\label{fig:BiasParam} Normalized gravitationally induced bias
	parameter as a function of $z$, where $b_{15}\equiv b(z=15)$, $z=15$
	is the typical condensation scale and $b={\rho_{B}/\rho_{DM}}=\Omega_{B_{0}}/\Omega_{DM_{0}}[\Omega_{DE_{0}}a^{\eta}+\Omega_{DM_{0}}]^{(\eta+3\omega_{DE})/\eta}.$}
\end{figure}

The deviation from the dust behaviour of DM due to the interactions with DE leads to evolution of the bias
parameter, $b=\rho_{B}/\rho_{DM}$, on cosmological
timescales [\cite{Berto07}a] (Fig. \ref{fig:BiasParam}). Other astrophysical
effects also affect the bias so the detection of this drift would
require statistics over different $z$ ranges. If we attribute the non-dust part of $\dot{\rho}_{DM}$ to an additional interaction, the differential of acceleration felt
by DM particles can be modeled  to be proportional to gravity, the Hubble time and the differential density flux [\cite{OBPedro07b}],
\begin{equation}
\frac{a_{Int.}}{a_{grav.}}=\delta\frac{\dot{\rho}_{DM}-\dot{\rho}_{DM}\Vert_{dust}}{\rho_{DM}H}=\delta\zeta~.\label{eq:RelInt}
\end{equation}

\begin{figure}[ht]
	\begin{center}\includegraphics[%
	  width=0.35\paperwidth,
	  height=0.67\textheight,
	  keepaspectratio]{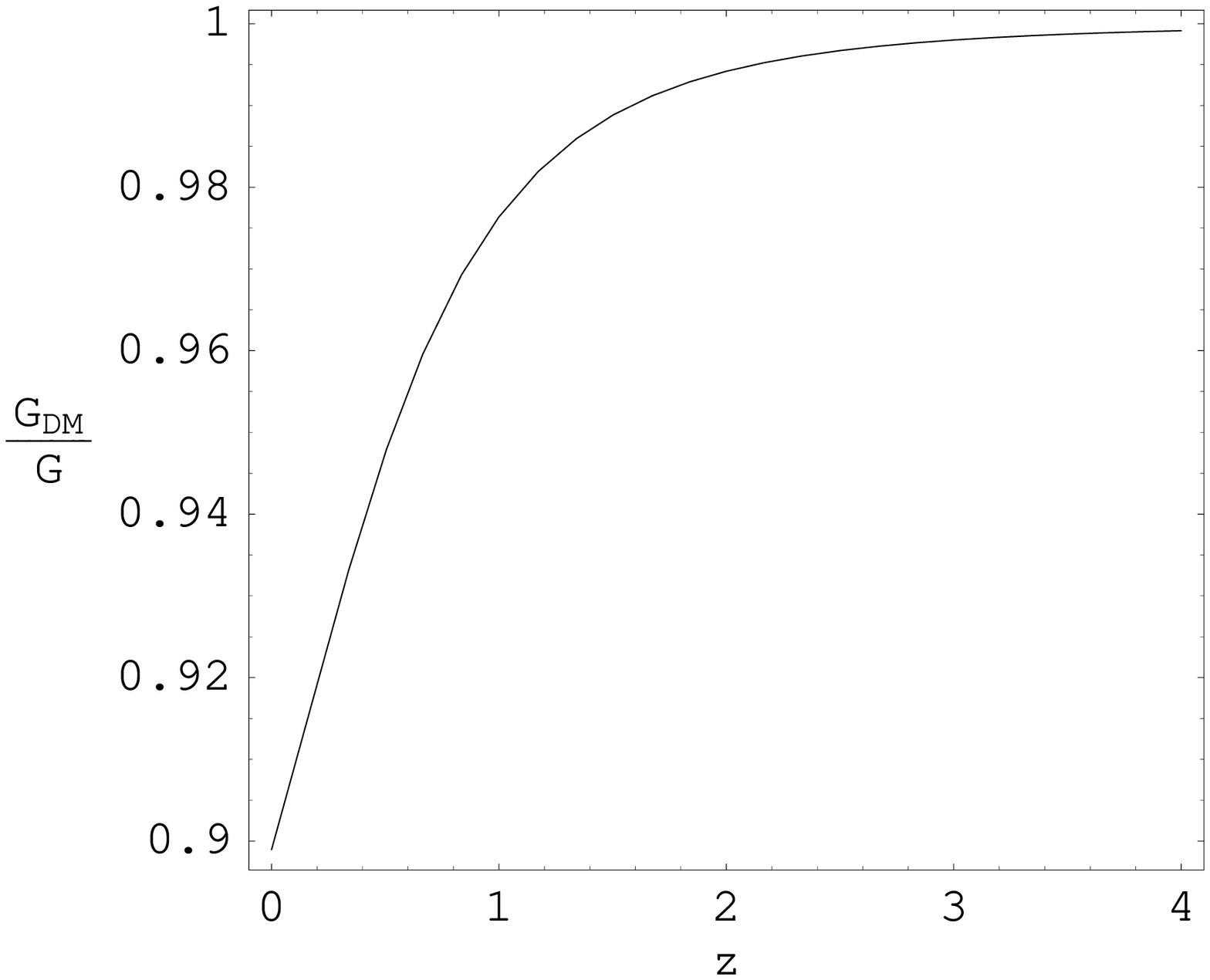}
	\end{center}

\caption{\label{fig:diffG} Evolution with redshift of the ratio of the
gravitational coupling for DM and baryons falling on a DM halo, using
the varying coupling model discussed in [\cite{OBPedro07b}].}
\end{figure}

Gravity is then Baryon/DM composition dependent and this effect depends only on cosmic time. We can now assign the time evolution to a varying
G, as seen in Fig. \ref{fig:diffG} [\cite{OBPedro07b}], to compare
with simulations on the fall on the Sagittarius Milky Way satellite on the DM of our galaxy which is consistent with $G_{DM}/G\leq1.1$ [\cite{KesdenKamion06}].

\noindent

\section{Conclusions}


Observations of cluster A586 [\cite{Cypriano:2005}] suggest evidence
of departure from virialization given that A586 is very spherical
and relaxed (from its mass distribution and Gyrs without mergers).
The generalized Layzer-Irvine equation allows to interpret this departure
as due to interaction with DE. We can therefore link the observed virialization to
interaction [\cite{Berto07}a,b] for two different interacting DE-DM  models:
an interacting quintessence with constant $\omega_{DE}$ {\tiny }[\cite{Amendola}]
and a Chaplygin gas with $\omega_{DE}=-1$ {\tiny }[\cite{Bento04}].

Based on the evidence of interaction, we argue that the Equivalence Principle should
be violated as seen through the bias parameter [\cite{Berto07}a] and on Baryon/DM asymmetric collapse [\cite{OBPedro07b}]. Our results are consistent with CMB, supernovae and Baryon acoustic oscillations as well as the simulation of the fall of the Sagittarius dwarf galaxy into our own. These results are quite encouraging and suggest that our method should be further employed into other cluster systems (see e.g. [\cite{Abdalla:2007rd}]).

\section*{Acknowledgement}


The work of MLeD is supported by FCT (Portugal), 
SFRH/BD/16630/2004, and hosted by J.P. Mimoso and CFTC, Lisbon University. The work of O.B. is partially supported by the FCT project POCI/FIS/56093/2004.


\end{document}